\documentclass[conference,letterpaper]{IEEEtran}
\usepackage{cite}
\usepackage[cmex10]{amsmath} % Use the [cmex10] option to ensure complicance
\usepackage{graphicx}
\usepackage{textcomp}
\usepackage[utf8]{inputenc}
\usepackage[T1]{fontenc}
\usepackage[ruled, vlined, nofillcomment, linesnumbered]{algorithm2e}
\usepackage{cancel}
\usepackage{fancyhdr}
\usepackage{txfonts}  %has \causeso and right
\usepackage{mathtools} %% double vertical bar
\usepackage{pgfplots} %% needed for confounder
\usepgflibrary{arrows}
\usepackage{subcaption}
\usepackage{bm} %% bold and italic
\usepackage{color}
\usepackage{arydshln}	% dotted lines for table
\usepackage{multirow}     % for tables
\usepgfplotslibrary{fillbetween}% for figures slope
\usepackage[pdfpagemode=None, colorlinks=true,  % url coloring
           linkcolor=black, urlcolor=black, citecolor=black, plainpages=false, 
           pdfpagelabels,unicode]{hyperref}
\RequirePackage{newclude}
\RequirePackage{pdfpages} 
\usepackage[noenumitem]{eda} % The EDA package, includes many default packages
\usepackage{tikz}
\usepackage{pifont} %%% used for cmark (crack)
\usepackage[english]{babel}
\usepackage{thmtools}   % for restating theorems
\usepackage{balance}

\graphicspath{ {./figs/} }

\ifpdf
\usetikzlibrary{external}
\fi

\newif\iflong
\longtrue % uncomment if you want the long version
\SetKwComment{tcpas}{\{}{\}}
\SetCommentSty{textnormal}
\SetArgSty{textnormal}
\SetKw{False}{false}
\SetKw{True}{true}
\SetKw{Null}{null}
\SetKwInOut{Output}{output}
\SetKwInOut{Input}{input}
\SetKw{AND}{and}
\SetKw{OR}{or}
\SetKw{Continue}{continue}

\makeatletter
\newcommand{\oset}[3][0ex]{%
  \mathrel{\mathop{#3}\limits^{
    \vbox to#1{\kern-2\ex@
    \hbox{$\scriptstyle#2$}\vss}}}}
\makeatother

\DeclarePairedDelimiterX{\infdivx}[2]{(}{)}{%
  #1\;\delimsize\|\;#2%
}

\DeclareMathAlphabet{\mathbcal}{OMS}{cmsy}{b}{n} %% mathbcal
\DeclareMathOperator*{\argmin}{\arg\!\min}
\DeclareMathOperator*{\argmax}{\arg\!\max}

\newcommand{\XtoY}{{X \rightarrow Y}\xspace}
\newcommand{\YtoX}{{Y \rightarrow X}\xspace}

\newcommand{\Models}{\ensuremath{\mathcal{M}}\xspace}

\newcommand{\model}{\ensuremath{M}}

\newcommand{\peq}{\ensuremath{\stackrel{+}{=}}}

\newcommand{\Indep}{\mathop{\perp\!\!\!\perp}\nolimits} 
\newcommand{\nIndep}{\mathop{\cancel\Indep}\nolimits}

\newcommand{\Pa}{\ensuremath{\text{Pa}}\xspace}

\newtheorem{theorem}{Theorem}

\newtheorem{lemma}{Lemma}

\newtheorem{postulate}{Postulate}

\newtheorem{definition}{Definition}

\tikzset{node/.style={black, draw=black, circle, minimum size=0.7cm, scale=0.8}} 
\tikzset{snode/.style={black, draw=black, fill=lightgray, circle, minimum size=0.7cm, scale=0.8}} 
\tikzset{dummy/.style={black, draw=black, circle, minimum size=0.55cm, scale=0.7}} 
\tikzset{latent/.style={black, draw=black, fill=lightgray, circle, minimum size=0.55cm, scale=0.7}} 

\tikzset{causes/.style={->,very thick,  color=black}} 
\tikzset{causesxor/.style={->,very thick, dashed, color=black}} 
\tikzset{causesxoro/.style={o->,very thick, dashed, color=black}} 
\tikzset{connected/.style={o-o,very thick, color=black}} 
\tikzset{connectedd/.style={o-o,very thick, dashed,  color=black}} 
\tikzset{ocauses/.style={o->,very thick,  color=black}} 
\tikzset{confounder/.style={<->,very thick,  color=black}} 
\tikzset{confounderxor/.style={<->,very thick, dashed, color=black}} 
\tikzset{confounderl/.style={<->,very thick,  color=black, bend left=45}} 
\tikzset{confounderr/.style={<->,very thick,  color=black, bend right=45}} 

\tikzstyle{flatlabel}  = [above, font = \tiny, inner sep = 1pt, text = black]
\tikzstyle{flatlabelb}  = [below, font = \tiny, inner sep = 1pt, text = black]
\tikzstyle{slopelabel}  = [sloped, above, font = \tiny, inner sep = 1pt, text = black]
\tikzstyle{slopelabelb}  = [sloped, below, font = \tiny, inner sep = 1pt, text = black]

\begin{document}
\title{Formally Justifying MDL-based Inference of Cause~and~Effect}

\author{\IEEEauthorblockN{Alexander Marx, Jilles Vreeken}
\IEEEauthorblockA{\textit{CISPA Helmholtz Center for Information Security} \\
Saarbr{\"{u}cken}, Germany \\
\{alexander.marx,jv\}@cispa.de}
}

\maketitle

\begin{abstract}
The algorithmic independence of conditionals, which postulates that the causal mechanism is algorithmically independent of the cause, has recently inspired many highly successful approaches to distinguish cause from effect given only observational data. Most popular among these is the idea to approximate algorithmic independence via two-part Minimum Description Length (MDL). Although intuitively sensible, the link between the original postulate and practical two-part MDL encodings is left vague. In this work, we close this gap by deriving a two-part formulation of this postulate, in terms of Kolmogorov complexity, which directly links to practical MDL encodings. To close the cycle, we prove that this formulation leads on expectation to the same inference result as the original postulate.
\end{abstract}

\begin{IEEEkeywords}
Kolmogorov complexity, MDL, Causality
\end{IEEEkeywords}

\section{Introduction}
Recovering causal networks from observational data is a challenging, yet important problem in science.
Traditional methods are only able to infer the true causal graph up to its Markov equivalence class, i.e., recover the correct undirected network and some of the edge directions~\cite{spirtes:00:book}. Inferring the network structure beyond such a Markov equivalence class boils down to identifying the causal direction in a bi-variate setting~\cite{peters:17:elements}. That is, given two dependent random variables $X$ and $Y$, we need to distinguish between the two Markov equivalent graphs $X \to Y$ an $X \leftarrow Y$. Recent results show that it is possible solve this task and thus infer all edge directions, if we are willing to make assumptions about the causal mechanism~\cite{peters:17:elements,shimizu:06:anm}.
One such assumption is the principle of independent mechanisms, which postulates that the causal mechanism (i.e. the conditional distribution $P_{\text{effect}|\text{cause}}$) is independent of the distribution of the cause $P_{\text{cause}}$~\cite{janzing:12:igci,sgouritsa:15:cure}. Vice versa, it is unlikely to find such an independence for the anti-causal direction. In practice, it is difficult to measure the independence between mechanism and cause directly. Thus, many recent approaches build upon an algorithmic version of this principle, the algorithmic independence of conditionals postulate~\cite{janzing:10:algomarkov}. This postulate states that if $X \to Y$ is the true graph, the Kolmogorov complexity of the factorization of $P_{XY}$ for the causal model, i.e., $K(P_X) + K(P_{Y|X})$, is shorter than for the anti-causal model $K(P_Y) + K(P_{X|Y})$.

Since Kolmogorov complexity is not computable, practical instantiations of the above postulate, build upon two-part Minimum Description Length (MDL)~\cite{rissanen:78:mdl}. Methods based on this idea are among the state-of-the-art for cause-effect inference on discrete~\cite{budhathoki:17:cisc,budhathoki:17:journalorigo}, continuous~\cite{kalainathan:19:generative,marx:19:sloppy,mitrovic:18:causalkernel,tagasovska:18:qccd} and mixed-type data~\cite{marx:18:crack}.
Despite the success of two-part MDL approaches in causal inference, the link between the theory of algorithmic independence and practical encodings is crude at best. While the postulate, formulated in terms of Kolmogorov complexity, only compares the complexities of the factorizations of the true distribution, e.g., $K(P_X) + K(P_{Y|X})$, the MDL formulations consider the data w.r.t. a model under consideration and the model itself. Simply put, the data part does not appear in the postulate itself, but only in the MDL approximation. Naively including the data breaks the asymmetry between the causal and anti-causal direction since information is symmetric. That is, given the shortest programs $x^*$ and $y^*$ to compute $x$ resp. $y$, $K(x) + K(y \mid x^*) = K(y) + K(x \mid y^*)$ holds up to an additive constant. Hence, it is crucial to analyze the link between both concepts more thoroughly, which is what we do in this paper. 

In particular, starting from the MDL framework, we first derive a two-part formulation of the algorithmic independence of conditionals in terms of Kolmogorov complexity. Then, we prove that our new formulation leads to an equivalent inference principle as the original postulate, which closes the cycle to connect both ideas. As a corollary, we investigate the implications of our findings for joint encodings, which encode data and model jointly. We emphasize that for such encodings is important to encode the model independent of the data, as otherwise the asymmetry between the description length of the causal and anti-causal model might vanish.

This paper is organized as follows. We discuss the principle of independent mechanisms in Sec.~\ref{sec:amc:im} and state the algorithmic independence of conditionals postulate in Sec.~\ref{sec:amc:amc}. Then we review two-part MDL approximations of this postulate in Sec.~\ref{sec:amc:mdl-solution} and provide our main result, the link between both inference criteria in Sec.~\ref{sec:amc:amc-data}. In Sec.~\ref{sec:amc:connection-joint} we discuss practical implications of our findings for joint descriptions and in Sec.~\ref{sec:amc:conclusion}, we round up with a conclusion.

\section{The Principle of Independent Mechanisms}
\label{sec:amc:im}

Before we introduce the algorithmic independence of conditionals postulate, we first discuss its statistical equivalent, the principle of independent mechanisms.

To illustrate the problem, we will first consider the smallest working example.
That is, assume we are given two dependent random variables $X$ and $Y$ for which we want to infer the underlying causal graph from observational data. That is, we assume the data is passively collected and represents an i.i.d. sample of joint distribution $P_{XY}$. According to \emph{Reichenbach's common cause principle}~\cite{reichenbach:56:direction}, the dependence between $X$ and $Y$ can be explained by three possible graphs.

\iflong
\begin{figure}[t]%
	\begin{minipage}[t]{.33\linewidth}
		\centering
		\includegraphics[]{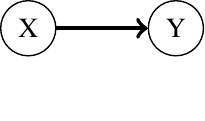}
		\subcaption{}
	\end{minipage}%
	\begin{minipage}[t]{.33\linewidth}
		\centering
		\includegraphics[]{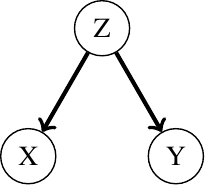}
		\subcaption{}
	\end{minipage}%
	\begin{minipage}[t]{.34\linewidth}
		\centering
		\includegraphics[]{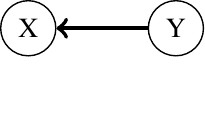}
		\subcaption{}
	\end{minipage}%
	\caption{Illustration of Reichenbach's Common Cause Principle: $X$ and $Y$ are two dependent random variables. Their dependence is either due to an unobserved confounder $Z$ (middle), or one of the two causes the other, i.e., either $X$ causes $Y$ (left) or $Y$ causes $X$ (right).}
	\label{fig:reichenbach}
\end{figure}
\else
\fi

\begin{definition}[Reichenbach's Common Cause Principle]
\label{def:acm:reichenbach}
If two random variables $X$ and $Y$ are statistically dependent ($X \nIndep Y$), then there exists a third variable $Z$ that causally influences both, that is $X \leftarrow Z \to Y$. As a special case, $Z$ may coincide with either $X$ or $Y$, which results in the causal graphs $X \to Y$ resp. $X \leftarrow Y$. \iflong See Figure~\ref{fig:reichenbach} for illustration.\else \fi
\end{definition}

It could also be that the dependence between $X$ and $Y$ is due to a combination of the above graphs, e.g. $X \to Y$, $X \leftarrow Z \to Y$. In this paper, however, we assume all causal relations to be acyclic and further assume \emph{causal sufficiency}, i.e., we observe all relevant variables. Thus, we only need to decide between $X \to Y$ and $X \leftarrow Y$, which is already a difficult problem since both these DAGs are Markov equivalent. Hence, it is not possible to tell apart both graphs if we rely on a constraint-based causal discovery approach~\cite{pearl:09:book}.

This is, however, not the end of the story. We can distinguish $X \to Y$ from $X \leftarrow Y$ if we additionally make assumptions about the generating mechanism. A very general such assumption, which has gained a lot of attention in recent years, is the \emph{principle of independent mechanisms}, which focuses on the possible factorizations of the joint distribution $P_{XY}$. In particular, we can write $P_{XY}$ as the product $P_XP_{Y|X}$ or $P_YP_{X|Y}$; but why does this help? Consider an example inspired by \cite[Ch.~2.1]{peters:17:elements}, in which the cause $X$ corresponds to the altitude and the effect $Y$ to the temperature as measured for different cities. Assume we consider a set of different cities from the same climate zone. If we observe the altitude for a random city, we will have a mechanism in mind to derive the corresponding temperature value (i.e. $P_{Y|X}$), which is independent of $P_X$. Further, we can make a thought experiment and think about how the temperature would change, if we were to change the altitude of the city, e.g., by magically lifting it into the air. Vice versa, it is hard to imagine that increasing or decreasing the temperature in a city, e.g., by putting on the heating system in every house, will change the altitude. In other words, the independence of the mechanism does not hold for the anti-causal direction: the mechanism $P_{X|Y}$ would need to take a rather particular form to be independent of $P_Y$, which only holds in specific settings, e.g., for a linear model with both the cause and additive noise being Gaussian distributed~\cite{peters:17:elements}. More generally, we can formulate the principle of independent mechanisms as follows~\cite{janzing:10:algomarkov,peters:17:elements}.

\begin{postulate}[Principle of Independent Mechanisms]
\label{pos:acm:im}
The causal generative process of a systems variables is composed of autonomous modules that do not inform or influence each other. In the probabilistic case, this means that the conditional distribution of each variable given its causes (i.e., its mechanism) does not inform or influence the conditional distributions of other variables.
\end{postulate}

Projected on our two-variable example, if we assume that the principle of independent mechanisms holds, we get that $P_X \Indep P_{Y|X}$, while the same does not necessarily hold for the factorization w.r.t. the anti-causal direction. There exist several approaches that aim at using this asymmetry to infer the causal direction between two random variables from observational data~\cite{janzing:12:igci,sgouritsa:15:cure}; in practice, however, it is difficult to precisely estimate this dependence. Therefore, we focus on the information theoretic variant of this principle~\cite{janzing:10:algomarkov}, which initiated the development of MDL-based estimators, which are among the state-of-the-art in the field~\cite{kaltenpoth:19:confounded,marx:19:sloppy,mitrovic:18:causalkernel,tagasovska:18:qccd}.

\section{The Algorithmic Model of Causality}
\label{sec:amc:amc}

In the following, we first introduce the algorithmic model of causality and the algorithmic independence of conditionals. After that, we state the commonly used two-part MDL approximation~\cite{budhathoki:17:journalorigo}.

Since the algorithmic model of causality is defined through Kolmogorov complexity, we will first provide a brief introduction to that topic. Intuitively, the Kolmogorov complexity of a finite binary string $x$ is the length of the shortest program that can output $x$ and then hold. The idea is that the more information or structure the string contains, the shorter is the program that can express it. Formally, we will in this paper refer to prefix Kolmogorov complexity~\cite{kolmogorov:65:information,vitanyi:93:book}.

\begin{definition}[Kolmogorov Complexity]
\label{def:kolmogorov-complexity}
The prefix Kolmogorov complexity of a finite binary string $x$ is the length of the shortest self-delimiting binary program $p^*$ for a universal prefix Turing machine $\mathcal{U}$ that generates $x$, and then halts, i.e.,
\[
K(x) = \min \{ |p| \mid p \in \{0,1\}^*, \mathcal{U}(p) = x \} \; .
\]
\end{definition}

To define conditional Kolmogorov complexity of a binary string $x$ given string $y$, we provide $y$ as input to the program that computes $x$ for free, that is
\[
K(x \mid y) = \min \{ |q| \mid q \in \{0,1\}^*, \mathcal{U}(y,q) = x \} \; .
\]
Finally, building upon the above definitions, we can define the algorithmic equivalent to mutual information, which we need below. For two binary strings $x$ and $y$, \emph{algorithmic mutual information}~\cite{chaitin:75:algindepcond} is defined as
\[
I_A(x ; y) := K(x) + K(y) - K(x,y) \; .
\]
Simply put, algorithmic mutual information is greater than zero if we can extract more structure by jointly compressing  $x$ and $y$ than with two individual programs. Equivalently, we can define $I_{A}$ as $K(x) + K(y \mid x^*)$, which holds up to an additive constant, which we denote by $\peq$. The $x^*$ in the conditional refers to the shortest program that describes $x$. Note that if we would instead use $x$ in the conditional, the equality would only hold up to a logarithmic constant dependent on $x$, which breaks the symmetry of the formulation~\cite{vitanyi:93:book}.

Now that we discussed the preliminary concepts, we can state the \emph{algorithmic model of causality} (AMC)~\cite{janzing:10:algomarkov}, which was proposed as an algorithmic equivalent of the statistical model of causality. Simply put, we can compute the value of $X$ with a program of constant complexity that gets as input the data over the parents of $X$ in the corresponding causal graph, and data w.r.t. an independent noise term.

\begin{postulate}[Algorithmic Model of Causality]
\label{pos:amc:amc}
Let $G$ be a DAG formalizing the causal structure among the strings $x_1, \dots, x_m$. Then every $x_i$ is computed by a program $q_i$ with length $\mathcal{O}(1)$ from its parents $\mathit{pa}_i$ and an additional input $n_i$. We write
\[
x_i = q_i( \mathit{pa}_i, n_i) \; ,
\]
meaning that the Turing machine computes $x_i$ from the input $\mathit{pa}_i, n_i$ using the additional program $q_i$ and halts. The inputs $n_i$ are jointly independent, i.e.,
\[
n_i \Indep n_1, \dots, n_{i-1}, n_{i+1}, \dots, n_m \; .
\]
\end{postulate}

Janzing and Sch{\"{o}}lkopf justified this model by showing that similar to the statistical model, we can also derive an algorithmic version of the causal Markov condition. That is, the \textit{algorithmic Markov condition} (AMC) states that the joint complexity over all nodes is given by the sum of the complexities of each individual node given the optimal compression of its parents
\begin{equation}
K(x_1, \dots, x_m) \peq \sum_{i=1}^m K(x_i \mid \mathit{pa}_i^*) \; .
\end{equation}
Due to the symmetry of information, i.e., $K(x) + K(y \mid x^*) \peq K(y) + K(x \mid y^*)$, the algorithmic Markov condition only allows for identifying the Markov equivalence class. To be able to distinguish between Markov equivalence classes, Janzing and Sch{\"{o}}lkopf~\cite{janzing:10:algomarkov} further postulated the algorithmic equivalent of the principle of independent mechanisms.

\begin{postulate}[Algorithmic Independence of Conditionals]
\label{pos:algindep}
Let $G$ be a causal DAG over a set of $m$ variables $\bm{X} = (X_1,\dots,X_m)$ with joint distribution $P_{\bm{X}}$, which is lower semi-computable, that is, $K(P_{\bm{X}}) < \infty$. A causal hypothesis is only acceptable if the shortest description of the joint distribution $P_{\bm{X}}$ is given by the concatenation of the shortest descriptions of the Markov kernels, i.e., 
\[
K(P_{X_1, \dots, X_m}) \peq \sum_{i = 1}^m K(P_{X_i | \Pa_i}) \; ,
\]
where $\Pa_i$ are the parents of $X_i$ in $G$. Equivalently,
\[
I_A(P_{X_1|\Pa_1}; \dots; P_{X_m|\Pa_m}) \peq 0 \; .
\]
\end{postulate}

If we apply the above postulate to the case where the true graph only consists of the edge $X \to Y$, we get that
\begin{equation}
\label{eq:indep-coditionals:pair}
I_A(P_X; P_{Y|X}) \peq 0 \; .
\end{equation}
Note that it is assumed that this independence only holds for the true causal direction. For additive noise models, for example, it has been shown that for the anti-causal direction we get a dependence~\cite{janzing:10:justifyanm}, that is, $I_A(P_Y; P_{X|Y}) \gg 0$. If we combine those results, we can derive an inference rule as follows. If $X \to Y$ is the true graph, then
\begin{equation}
\label{eq:amc:pairs}
K(P_X) + K(P_{Y|X}) \stackrel{+}{\le} K(P_Y) + K(P_{X|Y}) \; .
\end{equation}
In other words, we can infer the true causal direction by selecting the factorization with the smallest Kolmogorov complexity. To use this idea in practice, we first need to solve two problems. First, Kolmogorov complexity is not computable~\cite{vitanyi:93:book}, and second, we are not given the true distribution but just a limited number of data points. A principled way to solve at least the first part of the problem is to approximate Kolmogorov complexity via the Minimum Description Length principle, which we explain below.

\section{MDL as a Practical Solution}
\label{sec:amc:mdl-solution}

The Minimum Description Length (MDL) principle~\cite{grunwald:07:book,rissanen:78:mdl} is a practical variant of Kolmogorov Complexity, which approximates $K$ from above. Instead of considering all programs, we restrict ourselves to a certain model class $\Models$, for example, a certain class of parametric probability distributions. Given data $D$, which may represents a sample from a distribution $P_X$, our goal is to find that model $M^* \in \Models$, such that
\begin{equation}
\label{eq:mdl}
	M^* = \argmin_{M \in \Models} L(D \mid M) + L(M) \; ,
\end{equation}
where $L(M)$ is the length in bits needed to describe the model $M$ or identify $M$ within the model class $\Models$, and $L(D \mid M)$ is the length in bits of the description of data $D$ given $M$. This specific version of MDL is also referred to as two-part or crude MDL. 
As an example, consider that $\Models$ is the model class which refers to a multinomial distribution with $k$ categories. In this case $M$ refers to a specific $k$-dimensional parameter vector $\hat{\theta}$. Accordingly, $L(M)$ measures the complexity of identifying $M \in \Models$ or describing the parameter vector $\hat{\theta}$, and $L(D \mid M)$ refers to the negative log-likelihood of the data $D$, under the assumption that $D$ follows a multinomial distribution with parameter vector $\hat{\theta}$. That is, we encode a single data point $x$ as $- \log P(x \mid \hat{\theta})$. Note that by taking the negative logarithm with log base $2$, we arrive at a code length in bits. For more details to MDL, we refer to Gr{\"{u}}nwald~\cite{grunwald:07:book}.

The first idea on how the algorithmic independence of conditionals could lead to an MDL-based inference rule was sketched out by Janzing and Sch{\"{o}}lkopf~\cite{janzing:10:algomarkov}, however, they do neither instantiate nor evaluate this idea in practice. They suggest that, the probabilistic models $\hat{P}_X$ and $\hat{P}_{Y|X}$, which are learned from a finite number of observations, together define a joint distribution $\hat{P}_{\XtoY}$, which is not necessarily equal to the description of $\hat{P}_{\YtoX}$ in the inverse direction. As common in MDL, they first encode the complexity of the model, i.e., $\hat{P}_X$ and $\hat{P}_{Y|X}$, and then encode the data given the model as the negative log-likelihood w.r.t. $\hat{P}_{\XtoY}$ resp. $\hat{P}_{\YtoX}$ and select the direction with the smaller complexity as the causal one.

Budhathoki and Vreeken~\cite{budhathoki:17:journalorigo} suggested a more practical approximation of Eq.~\refeq{eq:amc:pairs} via two-part MDL as follows. For the causal direction, we define a model as $\model_{\XtoY}= (\model_{X}, \model_{Y\mid X})$ from the class $\Models_{\XtoY} = \Models_{X} \times \Models_{Y \mid X}$ that best describes the data over $Y$ by exploiting as much structure of $X$ as possible to save bits. By MDL, we identify the optimal model $\model_\XtoY \in \Models_{\XtoY}$ for data $(x^n,y^n)$ over $X$ and $Y$ as the one minimizing
\begin{equation}
\label{eq:lxtoy}
L_{\XtoY} := L(M_X) + L(x^n \mid M_X) + L(M_{Y|X}) + L(y^n \mid x^n, M_{Y|X}) \; .
\end{equation}
We can define $L_{\YtoX}$ analogously and infer $X \to Y$ if $L_{\XtoY} < L_{\YtoX}$, $X \leftarrow Y$ if $L_{\XtoY} > L_{\YtoX}$, and do not decide if both terms are equal. Consequently, to use this idea in practice, we need to define the model class. Budhathoki and Vreeken~\cite{budhathoki:17:journalorigo} implemented their idea for multivariate binary data and used binary trees as their models. Following this example, two-part MDL approaches have been developed for univariate discrete pairs~\cite{budhathoki:17:cisc,sy:20:causal}, univariate continuous random variables~\cite{kalainathan:19:generative,marx:17:slope,marx:19:sloppy,mitrovic:18:causalkernel,tagasovska:18:qccd} and multivariate mixed-type data~\cite{marx:18:crack}. Further, Kaltenpoth and Vreeken~\cite{kaltenpoth:19:confounded} adapted this idea to tell whether two random variables are causally related or whether they are likely to be confounded and Mian et al.~\cite{mian:20:globe} build upon a two-part MDL score to discover the complete causal DAG $G$ between a set of random variables.\!\footnote{There also exists approaches for continuous i.i.d. data~\cite{mooij:10:mml} and time series data~\cite{hlavavckova:20:grangerpoisson} based on the Minimum Message Length~\cite{wallace:68:mml}, however, these do not  directly build upon Eq.~\refeq{eq:amc:pairs}.}

Although these approaches perform well in practice, Eq.~\refeq{eq:amc:pairs} only considers the true distribution, while Eq.~\refeq{eq:lxtoy} is formulated via a two-part description of a model and the data given this model. In the following section, we will present our main result and formally analyze the connection between both inference rules. We bridge the gap between both variants by deriving a two-part variant of Eq.~\refeq{eq:amc:pairs}, in terms of Kolmogorov complexity, and show that on expectation both versions lead to the same inference.

\section{Linking Algorithmic Independence and Two-Part~Descriptions}
\label{sec:amc:amc-data}

Given an i.i.d. sample $x^n$ w.r.t. a distribution $P$, the shortest encoding of the data that is theoretically possible converges to the Shannon entropy
\[ 
H(P) = - \sum_x P(x) \log P(x) \; ,
\]
as proven by Shannon's source coding theorem~\cite{shannon:48:coding}. Hence, if $P$ is a computable distribution with parameter vector $\theta$, the sample estimate $\hat{\theta}$ will in the limit converge to the true parameter. Therefore, we could in the limit encode the data $x^n$ conditional on $P$ to arrive at the shortest code-length of the data given the model that describes $P$.
Thus, the shortest encoding for our causal setup can be achieved if the model class $\Models_X$ contains $P_X$ and similarly, $\Models_{Y|X}$ contains $P_{Y|X}$. Slightly abusing the notation, we define
\begin{equation}
\label{eq:lxtoystar}
L_{\XtoY}^* := L(P_X) + L(x^n \mid P_X) + L(P_{Y|X}) + L(y^n \mid x^n, P_{Y|X}) \; .
\end{equation}
The above equation already comes close to an MDL version of the algorithmic independence of conditionals, however, we still need to explain how the data encoded by the model fits into the equation.
To this end, we will show that the equivalent formulation of $L_{\XtoY}^*$ in terms of Kolmogorov complexity, i.e.,
\[
K_{\XtoY} := K(P_X) + K(x \mid P_X) + K(P_{Y|X}) + K(y \mid x, P_{Y|X}) \; ,
\]
is on expectation equal to $K(P_X) + K(P_{Y|X}) + H(P_{XY})$, where $H(P_{XY})$ relates to the Shannon entropy of the joint distribution $P_{XY}$. The analogoue holds for the anti-causal direction, that is, on expectation $K_{\YtoX}$ is equal to $K(P_Y) + K(P_{X|Y}) + H(P_{XY})$. Thus, assuming that the algorithmic independence of conditionals holds, we get that on expectation the inequality between cause and effect holds similar to Eq.~\ref{eq:amc:pairs}. That is,
\[
K_{\XtoY} \stackrel{+}{\le} K_{\YtoX} \; ,
\]
if $X \to Y$ is the true causal direction.

Before we prove this statement, we need to introduce a more general Lemma that links Kolmogorov complexity to Shannon entropy~\cite[Ch.~8.1]{vitanyi:93:book}.

\begin{lemma}
\label{le:equalitytoh}
Let $H(P)$ be the entropy of a computable probability distribution $P$ and $H(P) < \infty$. Then, 
\begin{equation}
\left| \, \sum_x P(x) K(x \mid P) - H(P) \, \right| \le \mathcal{O}(1),
\end{equation}
with a constant precision that is independent of $x$ and $P$.
\end{lemma}

Note that if we sum over $P(x)K(x)$ instead of $P(x)K(x \mid P)$, the inequality becomes less precise and only holds up to constant $c_P = K(P) + \mathcal{O}(1)$, which is dependent on $P$~\cite[Ch.~8.1]{vitanyi:93:book}. For conditional codes such as $K(y \mid x, P_{Y|X})$ assume that given input $x$ there exists an $\mathcal{O}(1)$ program that selects the correct probability table $P_{Y|X=x}$ from the auxiliary conditional probability table that is given as input. Based on these insights, we can derive of our main theorem.

\begin{theorem}
\label{th:main}
Given rational distribution $P_{XY}$ with finite support, for which all factorizations are lower semi-computable, i.e., $K(P_X) + K(P_{Y|X}) + K(P_Y) + K(P_{X|Y}) < \infty$, it holds that
\[
\sum_x \sum_y P_{XY}(x,y) \left( K(P_X){+}K(x \mid P_X){+}K(P_{Y|X}){+}K(y \mid x, P_{Y|X}) \right)
\]
is equal to $K(P_X) + K(P_{Y|X}) + H(P_{XY})$ up to an additive constant that is independent of $P_X$ and $P_{Y|X}$. Equivalently, $K(P_Y) + K(P_{X|Y}) + H(P_{XY})$ is equal to the expectation over $K(P_Y) + K(y \mid P_Y) + K(P_{X|Y}) + K(x \mid y, P_{X|Y})$ up to an additive constant independent of $P_Y$ and $P_{X|Y}$.
\end{theorem}
\begin{IEEEproof}
In the following, we prove the statement for the factorization $P_X P_{Y|X}$; the proof for the factorization $P_Y P_{X|Y}$ follows analogously. First, note that we can compute $P_X(x)$ as $P_X(x) = \sum_y P_{XY}(x,y)$. Thus, we can rewrite the first part as
\begin{align}
(\star_1) = &\sum_x \sum_y P_{XY}(x,y) \left( K(P_X) + K(x \mid P_X) \right) \\
= &\sum_x P_X(x) \left( K(P_X) + K(x \mid P_X) \right) \\
\peq \; &K(P_X) + H(P_X) \; .
\end{align}
To get from line $2$ to $3$, we apply Lemma~\ref{le:equalitytoh}.
Similarly, we can proceed with the second part
\begin{align}
(\star_2) = &\sum_x \sum_y P_{XY}(x,y) \left( K(P_{Y|X}) + K(y \mid x, P_{Y|X}) \right) \\
= &\sum_x P_X(x) \sum_y \frac{P_{XY}(x,y)}{P_X(x)} \left( K(P_{Y|X}) + K(y \mid x, P_{Y|X}) \right) \\
= &\; K(P_{Y|X}) + \sum_x P_X(x) \sum_y \frac{P_{XY}(x,y)}{P_X(x)} K(y \mid x, P_{Y|X}) \\
\peq &\; K(P_{Y|X}) + \sum_x P_X(x) \sum_y P_{Y|X=x}(y) K(y \mid P_{Y|X=x}) \\
\peq &\; K(P_{Y|X}) + \sum_x P_X(x) H(P_{Y|X=x}) \\
= &\; K(P_{Y|X}) + H(P_{Y|X}) \; .
\end{align}
Importantly, in the step from line $3$ to $4$, we assume that we can select the correct probability table $P_{Y|X=x}$ form inputs $P_{Y|X}$ and $x$ with an $\mathcal{O}(1)$ program.

If we combine $(\star_1)$ and $(\star_2)$, we get $K(P_X) + K(P_{Y|X}) + H(P_{XY})$ and obtain an equivalent result for the inverse direction due to the symmetry of the joint entropy.
\end{IEEEproof}

Although the theorem is only stated for two random variables, it is straight forward to extend it to the general formulation of the algorithmic independence of conditionals. In particular, we have
\[
\sum_{i = 1}^m \sum_{x_i} \sum_{\mathit{pa}_i} P_{X_i \Pa_i}(x_i, \mathit{pa}_i) \left( K(P_{X_i | \Pa_i}) + K(x_i \mid \mathit{pa}_i, P_{X_i | \Pa_i}) \right)
\]
is equal to $\sum_{i = 1}^m K(P_{X_i | \Pa_i}) + H(P_{X_1, \dots, X_m})$ up to a constant.

This concludes the main contribution of this paper. In the following section, we point out that the results of Theorem~\ref{th:main} do not necessarily hold for joint descriptions and discuss implications of these observations for practical MDL encodings.

\section{Connection to Joint Descriptions}
\label{sec:amc:connection-joint}

The optimization goal of a two-part encoding, e.g., two-part MDL, is also often formalized as finding that model $M^* \in \Models$, which mimimizes the \emph{joint} costs of data and model, that is, 
\[
M^* = \argmin_{M \in \Models} L(D,M) = L(M) + L(D \mid M) \; .
\]
Hence, intuitively we could rewrite $L_{\XtoY}$ as $L(x^n, M_X) + L(y^n, M_{Y|X} \mid x^n)$. The problem is, if we rigorously expand the second term, we need to encode the model given the data, i.e., 
\[
L(y^n, M_{Y|X} \mid x^n) = L(M_{Y|X} \mid x^n) + L(y^n \mid x^n, M_{Y|X}) \; .
\]
In terms of MDL, we can argue that the model is independent of the data $x^n$. In general, while technically possible, it is not common to encode a model conditioned on the data. Thus, we do not elaborate further on this ambiguity and jump into Kolmogorov land.

In particular, assume that $X \to Y$ is the true causal model. If we were to prove that $K(x,P_X) + K(y, P_{Y|X} \mid x)$ is on average equal to $K(P_X) + K(P_{Y|X}) + H(P_{XY})$, the proof would become slightly more involved. It is inevitable, however, that to split off $P_{Y|X}$ from $K(y, P_{Y|X} \mid x)$ we need to keep $x$ in the conditional term. That is, we arrive at the term $K(P_{Y|X} \mid x)$ and would need to argue that it is equal to $K(P_{Y|X})$. Since $P_X \Indep P_{Y|X}$ and $x$ is sampled from $P_X$, we can indeed conclude that $K(P_{Y|X} \mid x) = K(P_{Y|X}) + \mathcal{O}(1)$. For the anti-causal direction, this independence does not hold, i.e., $P_Y \nIndep P_{X|Y}$. Hence, $K(P_{X|Y} \mid y) \stackrel{+}{\le} K(P_{X|Y})$. Due to this asymmetry, we get that 
\begin{align}
&\sum_x \sum_y P_{XY}(x,y) (K(y,P_Y) + K(x,P_{X|Y} \mid y)) \\
&\stackrel{+}{\le} K(P_Y) + K(P_{X|Y}) + H(P_{XY}) \\
&\stackrel{+}{\ge} K(P_X) + K(P_{Y|X}) + H(P_{XY}) \; .
\end{align}
In other words, an approximation of this formulation does not allow us to distinguish $\XtoY$ from $\YtoX$, because we cannot guarantee that the inequality between the causal and anti-causal direction still holds. Thus, encodings that approximate the joint description with the goal to do causal inference should be designed with caution, as the description length of the conditional model should be independent of the data of the conditioning variable.

\section{Conclusion}
\label{sec:amc:conclusion}

In this paper, we focused on causal discovery from observational data. We revisited the idea of using two-part MDL encodings to approximate the algorithmic independence of conditionals postulate, which allows us to infer causal DAGs beyond Markov equivalence classes. Our main contribution is that we link this postulate to an alternative formulation, which considers both the complexity of the data conditioned on the distribution as well as the complexity of the distribution. Thus we arrive at a two-part description in terms of Kolmogorov complexity $K_{\XtoY}$, which has a one-to-one mapping to the proposed two-part MDL approximation $L_{\XtoY}$~\cite{budhathoki:17:journalorigo}. In addition, we prove that approximating $K_{\XtoY}$ is equivalent to approximating the algorithmic independence of conditionals, which closes the cycle. We further analyze the implications of our new formulation for joint descriptions of data and model and conclude that it is important that the model for the conditional distributions, e.g., $M_{Y|X}$ is independent of the data $x^n$, resp. that $M_{X|Y}$ is independent of $y^n$.

In a nutshell, drawing this connection is crucial to understand how two-part MDL approaches can be used to approximate the algorithmic independence of conditionals. We expect that these insights make MDL-based methods for causal inference more accessible to a broader audience.

\section*{Acknowledgment}
The authors would like to thank Bruno Bauwens and David Kaltenpoth for insightful discussions.

\balance
\bibliographystyle{IEEEtranS}
\bibliography{bib/abbreviations,bib/bib}

\end{document}